\documentclass[12pt]{article}

\usepackage{amssymb}
\usepackage{color,graphicx}
\usepackage{amsmath}

\usepackage{hyperref}

\begin{document}

\renewcommand{\abstractname}{\hfill}

\newpage
\pagenumbering{arabic}
\begin{center}
\LARGE \bf Black holes and particles with zero or negative energy
\end{center}

\vspace{0mm}
\begin{center}
\bf
A. A. Grib${}^*$\footnote{${}^*$Theoretical Physics and Astronomy Department
of the Herzen State Pedagogical University of Russia,
and
A.~Friedmann Laboratory for Theoretical Physics,
Saint Petersburg, Russia,

E-mail:\, andrei\_grib@mail.ru}
and Yu. V. Pavlov${}^{**}${}\footnote{${}^{**}$Institute of Problems in
Mechanical Engineering of Russian Academy of Sciences
and
N.I.~Lobachevsky Institute of Mathematics and Mechanics,
Kazan Federal University, Kazan, Russia,

E-mail:\, yuri.pavlov@mail.ru}
\end{center}

\begin{abstract}
    We study properties of particles with zero or negative energy and a nonzero
orbital angular momentum in the ergosphere of a rotating black hole.
    We show that the sign of the particle energy is uniquely determined
by the angular velocity of its rotation in the ergosphere.
    We give a simple proof of the fact that extreme black holes cannot exist.
    We investigate the question of the possibility of an unlimited energy
increase in the center-of-mass system of two colliding particles,
one or both of which have negative or zero energy.
\end{abstract}


\section{\large Introduction}

    As is known, in the ergosphere of rotating black holes described by
the Kerr metric, there are special geodesics along which particles with
a negative or zero energy (with a nonzero projection of angular momentum)
can move~\cite{MTW}--\cite{NovikovFrolov}.
    Such particles, of course, are not observed in the region outside
the ergosphere, where there are only particles with positive energy.

    Here, we systematically analyze properties of geodesics in
the ergosphere of a black hole and obtain bounds for the energy and
the projection of the orbital momentum of particles with any energy value.
    Based on these bounds, we prove the statement that there cannot be
an extreme black hole with a critical value of the intrinsic angular momentum
of the black hole rotation.
    We show that the geodesics of particles with zero energy in
the ergosphere of a black hole escape from under the gravitational radius and
then go back under the gravitational radius (just like the geodesics of
particles with negative energy as we previously showed~\cite{GribPavlovVert}).
    We obtain estimates for the angular velocity of particles with positive,
negative, and zero energy.

    In this paper, we use the system of units in which the gravitational
constant and the speed of light are equal to unity: $G=c=1$.

\vspace{4mm}
{\section{\large Geodesics in the Kerr metric}
\label{sec2}}

    The Kerr metric of a rotating black hole~\cite{Kerr63}
in the Boyer–Lindquist coordinates~\cite{BoyerLindquist67} has the form
    \begin{equation}
d s^2 = \frac{\rho^2 \Delta}{\Sigma^2}\, d t^2 -
\frac{\sin^2\! \theta}{\rho^2} \Sigma^2 \, ( d \varphi - \omega\, d t)^2
\label{Kerr}
- \frac{ \rho^2}{\Delta}\, d r^2 - \rho^2 d \theta^2 .
\end{equation}
    where
$$
\rho^2 = r^2 + a^2 \cos^2 \! \theta, \ \ \ \ \
\Delta = r^2 - 2 M r + a^2,
$$
$$
\Sigma^2 = (r^2 + a^2)^2 - a^2 \sin^2\! \theta\, \Delta , \ \ \ \
\omega = \frac{2 M r a}{\Sigma^2} ,
$$
    $M$ is the mass of the black hole, and $ aM $ is its angular momentum.
We assume that ${0 \le a \le M }$.
    The event horizon of the Kerr black hole is the surface given by
the equation
$$
r = r_H \equiv M + \sqrt{M^2 - a^2} .
$$
    The surface defined by the equation
$$
r = r_C \equiv M - \sqrt{M^2 - a^2} .
$$
    is called the Cauchy horizon.
    The surface of the static limit is defined by the formula
$$
r = r_1 \equiv M + \sqrt{M^2 - a^2 \cos^2 \theta} .
$$
    The region of space–time between the static limit and the event horizon
is called the ergosphere~\cite{MTW,NovikovFrolov}.
    The quantity
$$
S (r, \theta) = r^2 -2 M r + a^2 \cos^2 \! \theta
$$
    vanishes on the ergosphere boundary, and $ S (r, \theta) < 0$
inside the ergosphere.

    Using the relation
    \begin{equation} \label{SSrho}
S\, \Sigma^2 + 4 M^2 r^2 a^2 \sin^2\! \theta = \rho^4 \Delta ,
\end{equation}
    one can write the equations of geodesics for the Kerr metric~(\ref{Kerr})
(see~\cite{Chandrasekhar}, Sec.~62 or~\cite{NovikovFrolov}, Sec.~3.4.1)
in the form
    \begin{equation} \label{geodKerr1}
\rho^2 \frac{d t}{d \lambda } = \frac{1}{\Delta}
\left( \Sigma^2 E - 2 M r a J \right), \ \ \
\rho^2 \frac{d \varphi}{d \lambda } = \frac{1}{\Delta}
\left( 2 M r a E + \frac{S\, J}{\sin^2\! \theta} \right),
\end{equation}
    \begin{equation} \label{geodKerr3}
\rho^2 \frac{d r}{d \lambda} = \sigma_r \sqrt{R}, \ \ \ \ \
\rho^2 \frac{d \theta}{d \lambda} =\sigma_\theta \sqrt{\Theta},
\end{equation}
    where
    \begin{equation} \label{geodR}
R = \Sigma^2 E^2 - \frac{S\, J^2}{\sin^2 \theta } - 4 M r a E J -
\Delta \left[ m^2 \rho^2 + \Theta \right],
\end{equation}
    \begin{equation} \label{geodTh}
\Theta = Q - \cos^2 \! \theta \left[ a^2 ( m^2 - E^2) +
\frac{J^2}{\sin^2 \! \theta} \right].
\end{equation}
    Here $E={\rm const}$ is the energy of a moving particle
(called the energy at infinity in~\cite{MTW}),
$J$~is the conserved projection of the particle angular momentum on
the black hole rotation axis, $m$~is the rest mass of the moving particle,
$\lambda $~is an affine parameter along the geodesic
    ($\lambda = \tau /m$ for a particle with $m \ne 0$, where
$\tau$ is its proper time), and $Q$ is the Carter constant
($Q=0$ when moving in the equatorial plane $\theta = \pi/2$).
    The constants $\sigma_r, \sigma_\theta = \pm 1$ determine the direction
of motion with respect to the coordinates $r$ and $\theta$.

    Geodesic equations~(\ref{geodKerr1}), (\ref{geodKerr3})
coincide with the Euler–Lagrange equations for the Lagrangian
    \begin{equation}
{L}  = \frac{g_{ik}}{2} \, \frac{d x^i}{d \lambda} \frac{d x^k}{d \lambda} .
\label{Lgeod}
\end{equation}
    The corresponding generalized momenta are equal to
    \begin{equation}
p_i  = \frac{\partial L}{\partial \left( \frac{d x^i}{ d \lambda } \right) }
= g_{ik} \frac{d x^k}{d \lambda } ,
\label{Lpdef}
\end{equation}
    \begin{equation}
p_t = E , \ \ \ p_r = - \sigma_r \frac{\sqrt{R}}{\Delta} , \ \ \
p_\theta = - \sigma_\theta \sqrt{\Theta} , \ \ \ p_\varphi = - J .
\label{pkovi}
\end{equation}
    It is easy to verify by direct calculation that
$$
p_i\, p_k\, g^{ik} = m^2 .
$$
    We note that the signs of the momentum components in~(\ref{pkovi})
correspond to the signs of the covariant components of the four-momentum
in~\cite{LL_II} (see Sec.\,9): $p_i=(E, - \mathbf{p}$).
    The momenta $p_t$ and $p_\varphi $ are constant on geodesics,
i.e. $E$ and $J$ are constant, which follows from the fact that the components
of Kerr metric~(\ref{Kerr}) are independent of the time~$t$ and angular
coordinate~$\varphi$.
    The fact that the parameter~$E$ preserved on geodesics is interpreted as
the particle energy can also be obtained based on the metric energy-momentum
tensor of the classical action of a point particle~\cite{GribPavlov2010NE}.

    As follows from~(\ref{geodKerr3}), the parameters characterizing any
geodesic should satisfy the conditions
    \begin{equation} \label{ThB0}
R \ge 0, \ \ \ \ \ \Theta \ge 0 .
\end{equation}
    For a geodesic corresponding to the trajectory of a moving test particle
outside the event horizon, the condition of motion ``forward in time''
must be satisfied:
    \begin{equation} \label{ThB0t}
d t / d \lambda > 0 .
\end{equation}
    Conditions~(\ref{ThB0}) and (\ref{ThB0t}) lead to the following
inequalities for the possible energy values~$ E $ and the angular
momentum projection~$J$ of a test particle at a point with the coordinates
$(r, \theta)$ with a fixed value of~$\Theta \ge 0$~\cite{GribPavlov2013}.

    $\bullet $
    Outside the ergosphere $ S(r, \theta) >0 $,
    \begin{equation} \label{EvErg}
E \ge \frac{1}{\rho^2} \sqrt{(m^2 \rho^2 + \Theta) S},
\ \  \ J \in \left[ J_{-} (r,\theta), \ J_{+} (r,\theta) \right],
\end{equation}
    \begin{equation}
J_{\pm} (r,\theta) = \frac{\sin \theta}{S} \left[ - 2 r M a E \sin \theta \pm
\sqrt{ \Delta \left( \rho^4 E^2 - (m^2 \rho^2 + \Theta) S \right)} \right].
\label{Jpm}
\end{equation}

    $\bullet $
    On the boundary of the ergosphere (for $\theta \ne 0, \pi$)
    \begin{equation} \label{rEgErg}
r = r_1(\theta) \ \ \ \Rightarrow \ \ \ E \ge 0,
\end{equation}
    \begin{equation} \label{JgErg}
J \le E \left[ \frac{M r_1(\theta) }{a} + a \sin^2 \! \theta \left(
1 - \frac{m^2}{2 E^2} - \frac{\Theta}{4 M r_1(\theta) E^2} \right) \right],
\end{equation}
    with the possible value $E=0$ if $m=0$ and $\Theta=0$,
in which case any value $J <0$ is allowed.

    $\bullet $
    Inside the ergosphere for $ r_H < r < r_1(\theta) $ and $S < 0$,
    \begin{equation} \label{lHmdd}
J \le \frac{\sin \theta}{- S} \left[ 2 r M a E \sin \theta -
\sqrt{ \Delta \left( \rho^4 E^2 - (m^2 \rho^2 + \Theta) S \right)} \right],
\end{equation}
    and the particle energy, as is known, can have any value,
both positive and negative.

    As can be seen from inequalities~(\ref{JgErg}) and (\ref{lHmdd}),
the angular momentum projection of a particle moving along a geodesic at
the boundary and inside the ergosphere can be negative and arbitrarily
large in absolute value for a fixed energy value.
    This property, found in~\cite{GribPavlov2013,GribPavlov2012} for
the Kerr metric, holds in the ergosphere of any black hole with an axially
symmetric metric as was later shown in~\cite{Zaslavskii13c}.

    If the particle angular momentum projection $J$ and a value $\Theta \ge 0$
are given, then from conditions~(\ref{ThB0}) and (\ref{ThB0t}), we obtain
    \begin{equation}
E \ge \frac{1}{\Sigma^2} \left[ 2 M r a J + \sqrt{ \Delta \left(
\frac{ \rho^4 J^2 }{\sin^2\! \theta } + \left( m^2 \rho^2 + \Theta \right)
\Sigma^2 \right) } \right]
\label{EVnHS}
\end{equation}
    anywhere outside the horizon.
    The lower bound of the energy values here corresponds to $R=0$.
    It can be seen from this inequality that the energy in the ergosphere
can be negative only if the particle angular momentum projection is negative.

\vspace{4mm}
{\section{\large  Effect of incident particles on the black hole rotation}
\label{secBReac}}

    As $r$ tends to the horizon $ r_H$ (for $\theta \ne 0, \pi$),
from~(\ref{lHmdd}) and~(\ref{EVnHS}), we obtain
    \begin{equation} \label{JgEH}
J \le J_H = \frac{ 2 M r_H E}{a}, \ \ \ \
E \ge \frac{a J}{ 2 M r_H}.
\end{equation}
    Hence, $J_H$ is the upper bound of values of the angular momentum
projection of a particle with the energy~$E$ at the black hole event horizon.

    We use the first inequality in~(\ref{JgEH}) to estimate the effect of
incident particles on the dimensionless angular momentum $A = a/M$
of the black hole.
    From the energy and momentum conservation laws, we obtain the formula
for the dimensionless angular momentum of the black hole after a particle
falls through the event horizon:
    \begin{equation} \label{AP1}
A' = \frac{a M + J}{(M + E)^2}.
\end{equation}
    Then
    \begin{equation} \label{AP2}
A' - A = \frac{ J - A E^2 - 2 A M E}{(M+E)^2 } \le
\frac{ J_H - A E^2 - 2 A M E}{(M+E)^2 } =
\frac{2 E \left( M \displaystyle \frac{r_H}{a} - a \right) -
\displaystyle E^2 \frac{a}{M}}{(M+E)^2 }.
\end{equation}
    Setting $A=1$ in this relation and taking into account that $ r_H = a = M$
in this case, we obtain the proof of the following statement.\\

    {\bf Statement 1}.
\emph{The fall of a particle into an extreme black hole with $A=1$ leads to
a decrease in its dimensionless angular momentum.
The black hole becomes nonextreme.}

\vspace{9pt}
    This indicates that extreme black holes cannot exist in nature.
    We note that the estimate in~\cite{Thorne74} for the limit angular momentum
of a black hole attainable with accretion of matter on it is $a=0.998M$.

    Taking into account that $ J < 0 $ at zero energy and using the left-hand
side of inequality~(\ref{AP2}) for zero energy and the right-hand side
of~(\ref{AP2}) for negative energies, we obtain the following statement.\\

    {\bf Statement 2}.
{\it The fall of a particle with zero or negative energy into a black hole
with $A \le 1$ leads to a decrease in its dimensionless angular momentum.}

\vspace{9pt}
    For test particles ($|E|/M \ll 1$) with $J \approx J_H$, we have
    \begin{equation} \label{AP2d}
A' - A \approx \frac{2 E \left( M \displaystyle \frac{r_H}{a} - a \right) -
\displaystyle E^2 \frac{a}{M}}{(M+E)^2 } \approx
\frac{2 E \left( M \displaystyle \frac{r_H}{a} - a \right)}{(M+E)^2 }.
\end{equation}
    Therefore, we have the following statement.\\

    {\bf Statement 3}.
{\it The fall of a test particle with positive energy and an angular momentum
projection close to the maximum value $J_H$ into a nonextreme black hole
with~$A<1$ leads to an increase in its dimensionless angular momentum.}

\vspace{9pt}
    We here note that for nonextreme black holes, the admissible values of~$J$
in the neighborhood outside the horizon are always strictly less than~$J_H$
(see, e.g., \cite{GribPavlov2011b}).

    In the case of particles freely falling into a nonextreme black hole from
infinity, we prove the following statement.\\

    {\bf Statement 4}.
{\it The dimensionless angular momentum of a nonextreme black hole can always
be increased by the free fall from infinity of specially selected particles.}

\vspace{9pt}
    {\bf Proof.}
    A simple analysis of the geodesic equations allows verifying that for
nonrelativistic particles at infinity with $E=m$, the condition for a fall
into a black hole from infinity in the equatorial plane is the inequality
    \begin{equation} \label{nerJ}
-2 (1+ \sqrt{1+A} ) \le \frac{J}{mM} \le 2 (1+ \sqrt{1-A} ).
\end{equation}
    In the fall of a test particle with $E=m \ll M$
and $J= 2 (1+ \sqrt{1-A} ) mM $, we have
    \begin{equation} \label{AP2dd}
A' - A = \frac{2 (1+ \sqrt{1-A} ) m M - A m^2 - 2 A mM}{(M+m)^2 }
\approx \frac{2 mM}{(M+m)^2 } \left( 1 - A + \sqrt{1-A} \right) > 0,
\end{equation}
    which proves the statement.

\vspace{9pt}
    We note that because of the simplicity of the proofs of our statements,
there is no need to use the laws of black hole mechanics formulated in
the famous paper~\cite{HawkingBarCar73} by analogy with thermodynamics.

\vspace{4mm}
{\section{\large  Particles with zero energy in the Kerr metric}
\label{secNulEn}}

    The study of the properties of a zero-energy particle is practically
absent from the literature on black holes. We consider the features of
geodesics for such particles.

    The value $E=0$ is possible at the boundary and inside the ergosphere.
    From the condition $\Theta \ge 0$ (see~(\ref{ThB0}))
and expression~(\ref{geodTh}), we obtain
    \begin{equation}
E =0 \ \ \Rightarrow \ \ Q \ge 0 ,
\label{E0Q}
\end{equation}
    i.e., {\it the Carter constant is nonnegative for test particles with
zero energy.}

    From condition~(\ref{ThB0t}) and expression~(\ref{geodKerr1})
for $dt / d \lambda $, we obtain
    \begin{equation}
E =0, \ \ \ \rho^2 \frac{d t }{d \lambda } = -
\frac{2 r M a J }{\Delta} > 0 \ \ \ \Rightarrow \ \ J < 0 ,
\label{E0Qdtl}
\end{equation}
    i.e., {\it the angular momentum projection of zero-energy particles
is negative.}
    From inequality~(\ref{lHmdd}) for zero-energy particles inside
the ergosphere, we have
    \begin{equation}
E =0 \ \ \ \Rightarrow \ \ \
J \le - \sin \theta \sqrt{\frac{\Delta \left(m^2 \rho^2 + \Theta \right)}{-S}}.
\label{E0Jtoch}
\end{equation}
    In the case of zero energy of a particle,
    \begin{equation}
E =0 \ \ \Rightarrow \ \ R = - \Delta ( m^2 \rho^2 + \Theta ) -
\frac{J^2}{\sin^2 \theta } S(r, \theta) .
\label{E0R}
\end{equation}
    Therefore, {\it the upper point of the trajectory of a massive particle
with zero energy is inside the ergosphere, $( r < r_1(\theta ))$.}

    Motion along the coordinate $\theta$  can continue as long as
$\Theta (\theta) $ does not vanish.
    Therefore, in the case $m=0$, we find from~(\ref{E0R}) that
    {\it the upper point of the trajectory of a massless zero-energy particles
is on the ergosphere boundary, $ r = r_1(\theta )$.}

    It follows from the geodesic equation at zero energy that
    \begin{equation}
\frac{d \varphi}{d t}  = \frac{- S(r)}{2 r M a \sin^2 \! \theta}
= \frac{r^2 - 2 r M + a^2 \cos^2\! \theta}{ - 2 r M a \sin^2 \! \theta}.
\label{omE0}
\end{equation}
    Therefore, {\it the angular velocity of any zero-energy particle is
independent of the mass and momentum of the particle and
is given by~(\ref{omE0}).}
    If a zero-energy particle reaches the ergosphere boundary
(hence, $m=0$ necessarily), then $S(r)=0$ and
    \begin{equation}
r\to r_1(\theta) \ \ \Rightarrow \ \ \frac{d \varphi}{d t} \to 0.
\label{omE0lim}
\end{equation}

    For the radial velocity of a particle with zero energy, we have
    \begin{equation}
E =0 \ \ \Rightarrow \ \ \frac{d r}{d t} = \frac{ \sigma_r \Delta }{2 M r a}
\sqrt{ \frac{-S}{\sin^2\! \theta} - \Delta \frac{ m^2 \rho^2 + \Theta}{J^2} }.
\label{rtE0}
\end{equation}
    Therefore, the radial velocity, generally speaking, depends on the angular
momentum.
    Namely, in the case $m^2 \rho^2 + \Theta >0$ at a point with a given $r$
in the ergosphere, the larger   $|J|$ is, the larger the radial velocity in
absolute value, and it can range from
$$
\frac{d r}{d t} =0 \ \ {\rm at} \ \
J= - \sin \theta \sqrt { \frac{\Delta (m^2 \rho^2 + \Theta )}{- S} }
$$
    to
$$
\frac{d r}{d t} = \frac{\sigma_r \Delta \sqrt{-S} }{2 M r a \sin \theta}
\ \ {\rm at} \ \ J \to - \infty .
$$

\begin{figure}[ht]
\centering
\includegraphics[width=50mm]{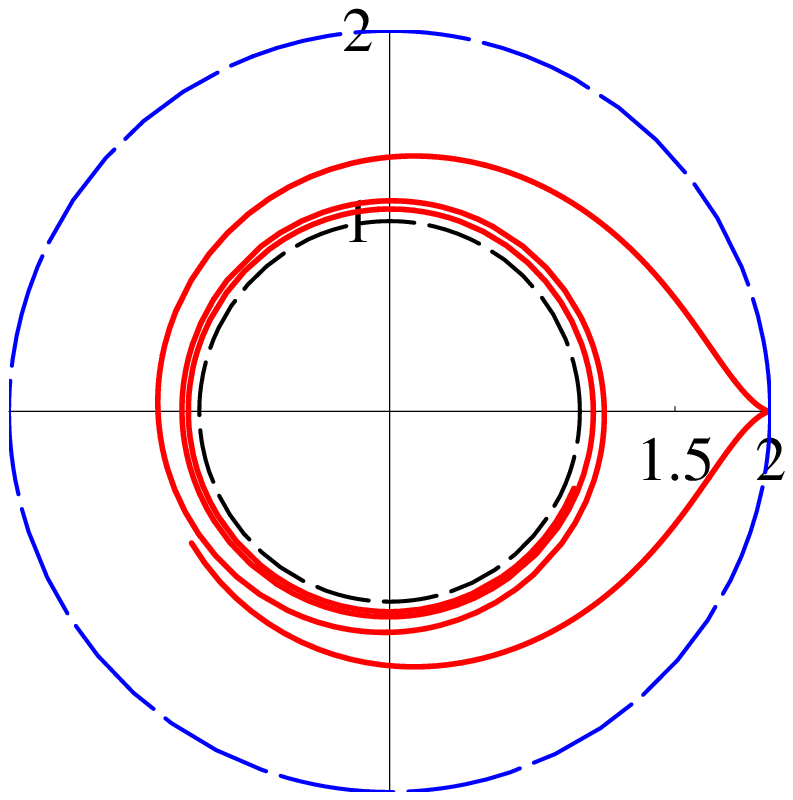} \hspace{14mm}
\includegraphics[width=50mm]{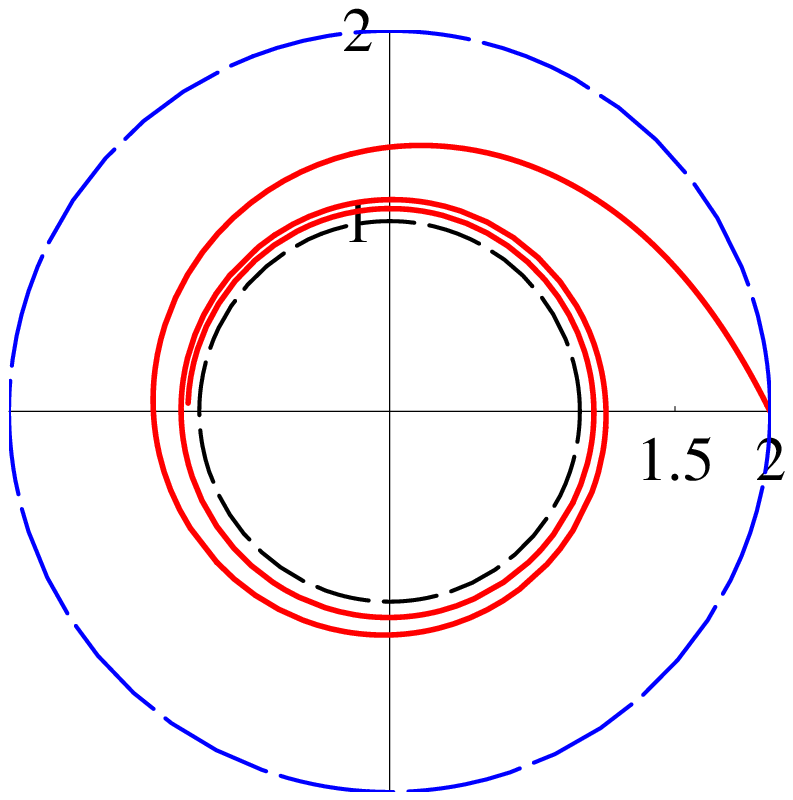} \\
\hspace{22mm} (a) \hspace{57mm} (b)\hspace{23mm}
\caption{The trajectory of photons in the ergosphere of a black hole
with $a=M$ in the cases where (a) $E=0$ and (b) $E>0$ and $J = M E $.}
\label{Fig}
\end{figure}

    The trajectory of photons in the equatorial plane ($m=0$, $ \Theta =0$)
is independent of the angular momentum, as can be seen from
Eqs.~(\ref{geodKerr1}) and (\ref{geodKerr3}).
    In polar coordinates, the equation of this trajectory can be written in
elementary functions and has the simplest form at $a=M$:
    \begin{equation}
\varphi(r) - \varphi(2M) = \pm \left( \arcsin \left( \frac{r}{M} - 1 \right) +
\frac{\sqrt{(2M - r) r}}{r-M} - \frac{\pi}{2}\right).
\label{pre0}
\end{equation}
    We show part of the trajectory in Fig.~\ref{Fig}\,a, where we use
the radial coordinate $r/M$.
    In Fig.~\ref{Fig}\,b, we show the trajectory of a photon moving in
the equatorial plane along a geodesic with positive energy and the orbital
angular momentum projection $J=ME$.
    The equation of this geodesic at $a=M$ has the form
    \begin{equation}
\varphi(r) - \varphi(2M) = \pm \frac{2M - r}{ r - M }.
\label{prep}
\end{equation}
    The presented trajectories, like trajectories of any particles crossing
the event horizon of the black hole, wrap around the horizon infinitely
many times.
    Indeed, it follows from geodesics equations~(\ref{geodKerr1}) and
(\ref{geodKerr3}) that
    \begin{equation}
\frac{d \varphi}{d r} \sim \frac{1}{\Delta} = \frac{1}{(r-r_H)(r-r_C)}
\label{PhirH}
\end{equation}
    near the horizon.
    We note that except in the case where $a=M$ and $J=J_H > 0 $, a massive
particle makes an infinite number of turns in a finite proper time!

    We previously showed~\cite{GribPavlovVert} that in the ergosphere,
the geodesics of particles with negative energies begin and end on
the surface~$r_H$.
    We prove here that
 {\it in the ergosphere, geodesics of zero-energy particles also
begin and end on the surface~$r_H$.}
    For this, we define the effective potential by the formula
    \begin{equation} \label{Leff}
V_{\rm eff} = - \frac{R}{2 \rho^4}.
\end{equation}
    In accordance with Eq.~(\ref{geodKerr3}), we have
    \begin{equation} \label{LeffUR}
\frac{1}{2} \left( \frac{d r}{d \lambda} \right)^{\!2} + V_{\rm eff}=0 ,
\ \ \ \ \frac{d^2 r}{d \lambda^2} = - \frac{d V_{\rm eff}}{d r} .
\end{equation}
    The necessary condition of existence of orbits with a constant $r$
(spherical orbits) can be written as
    \begin{equation} \label{LeffCucl}
V_{\rm eff}=0, \ \ \ \ \frac{d V_{\rm eff}}{d r} =0\,.
\end{equation}
    To prove the above statement, it suffices to prove
(see~\cite{GribPavlovVert}) that
    \begin{equation} \label{LeffdVefg0}
E=0, \ \ \ r > r_H, \ \ \ V_{\rm eff}(r)=0 \ \ \Rightarrow
\ \ V^{\, \prime}_{\rm eff}(r) > 0 .
\end{equation}
    Differentiating~(\ref{E0R}), we obtains
    \begin{equation}
E =0 \ \ \Rightarrow \ \ R'(r) = - 2 \left[ (r-M) \left(
\frac{J^2}{\sin^2 \theta} + m^2 \rho^2 + \Theta \right) + m^2 r \Delta \right],
\label{E0dR}
\end{equation}
    whence it clearly follows that condition~(\ref{LeffdVefg0}) holds.
    This proves our statement and, in particular, shows the absence of
spherical orbits for zero-energy particles.

    We note that the finiteness of the proper time of motion of zero-energy
particles in the ergosphere (or the affine parameter $\lambda$ for photons)
follows from~(\ref{LeffdVefg0}), as shown in~\cite{GribPavlovVert}.

    As can be seen from Fig.~\ref{Fig}, there are trajectories of photons
with positive energy (and also, as can be seen from similar figures, with
negative energy) that visually differ little from the trajectory of
a zero-energy photon.
    We further consider the question of which simple properties of the particle
motion in the ergosphere allow distinguishing particles with negative, zero,
and positive energy.

\vspace{4mm}
{\section{\large The angular velocity of particles in the ergosphere}
\label{seOmeg}}

    We find the constraints on the angular velocity of particles in
the ergosphere from the condition $ds^2 \ge 0$.
    We have
    \begin{equation}
g_{00}\, d t^2 + 2 g_{0 \varphi}\, d t\, d \varphi +
g_{\varphi \varphi}\, d \varphi^2 \ge 0,
\label{omN1}
\end{equation}
    and the angular velocity $\Omega = d \varphi / d t $
of any particle satisfies the constraints~\cite{LPPT}
    \begin{equation}
\Omega_1 \le \Omega \le \Omega_2, \ \ \ \
\Omega_{1,2} = \frac{g_{0 \varphi} \mp \sqrt{ g_{0 \varphi}^{\,2} -
g_{00} g_{\varphi \varphi} }}{- g_{\varphi \varphi} }.
\label{omN2}
\end{equation}
    On the ergosphere boundary, $g_{00} =0 $ and $\Omega_1 =0$.
    Inside the ergosphere, $g_{00} < 0 $, $\Omega_{1,2} > 0$,
and all the particles move in the direction of the black hole
rotation~\cite{MTW}--\cite{NovikovFrolov}.
    In approaching the event horizon,
    \begin{equation}
\lim \limits_{\ \  r \to r_H} \Omega_1 (r) = \! \!
\lim \limits_{\ \  r \to r_H} \Omega_2 (r) = \omega_{\rm Bh}
= \frac{a}{2 M r_H}.
\label{omN3}
\end{equation}
    The value $\omega_{\rm Bh}$ is called the angular velocity of
the black hole rotation.

\begin{figure}[ht]
\centering
\includegraphics[width=60mm]{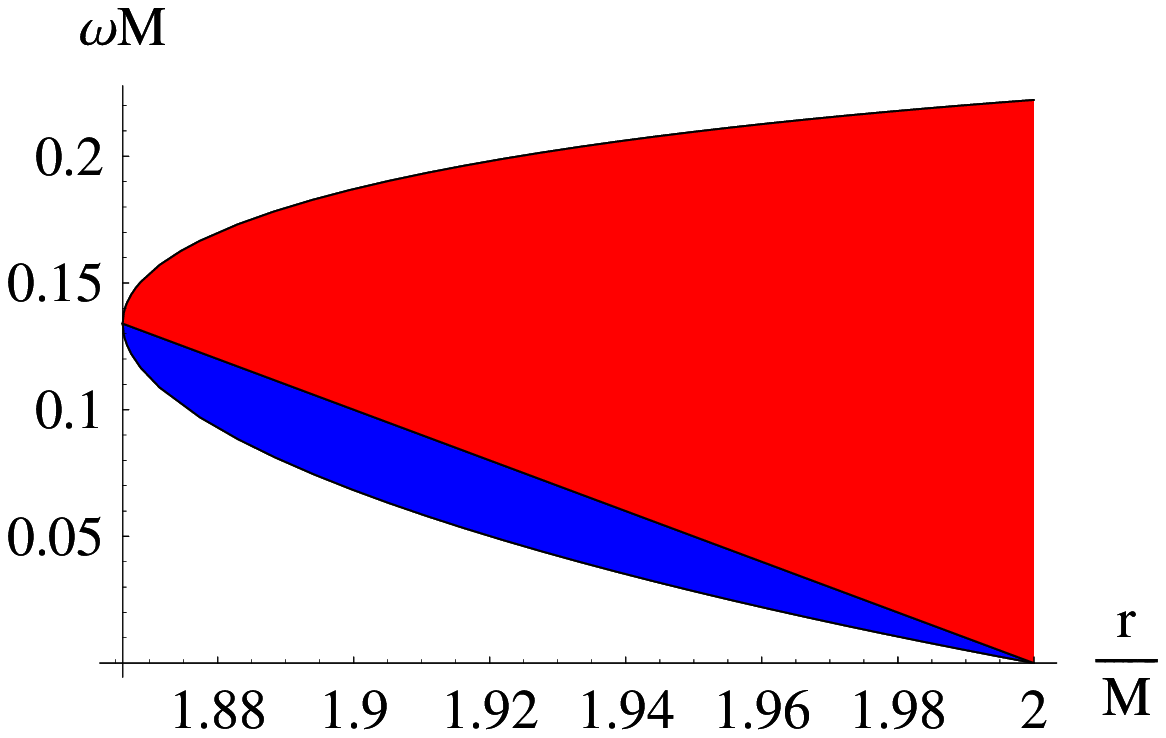} \ \ \ \
\includegraphics[width=60mm]{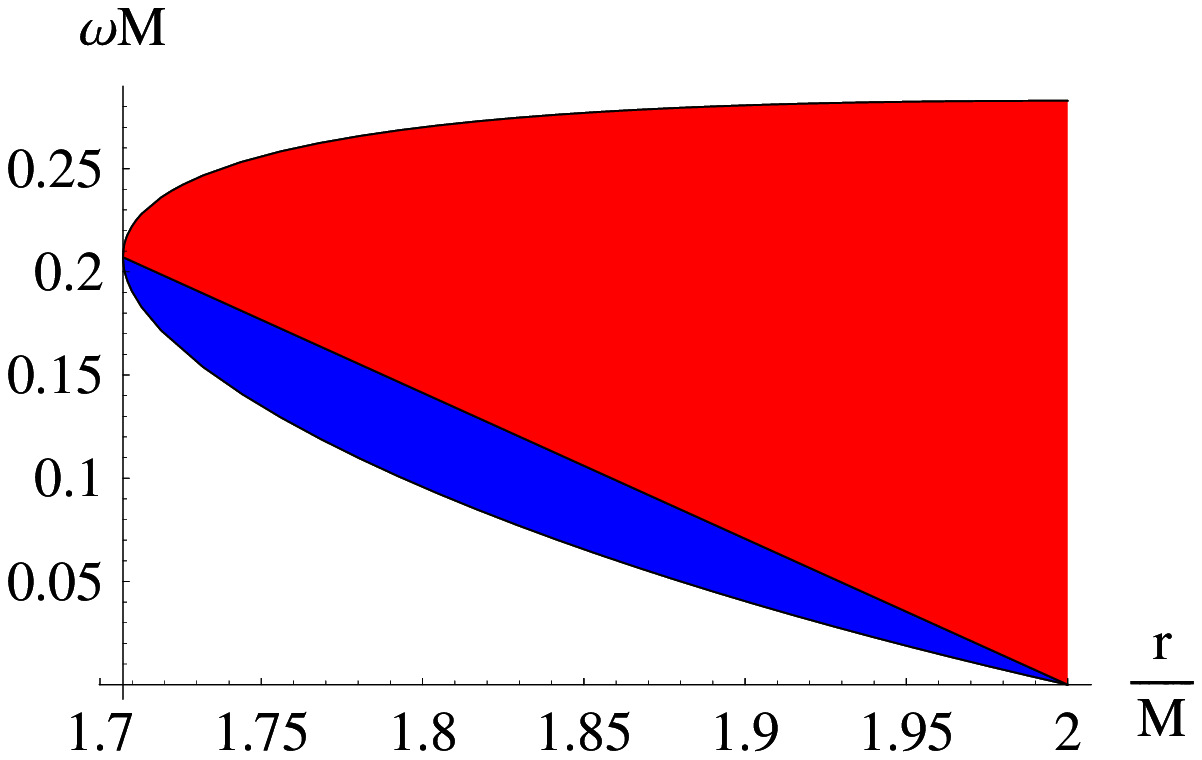} \\
\hspace{0mm}(a) \hspace{61mm} (b) \\
\includegraphics[width=60mm]{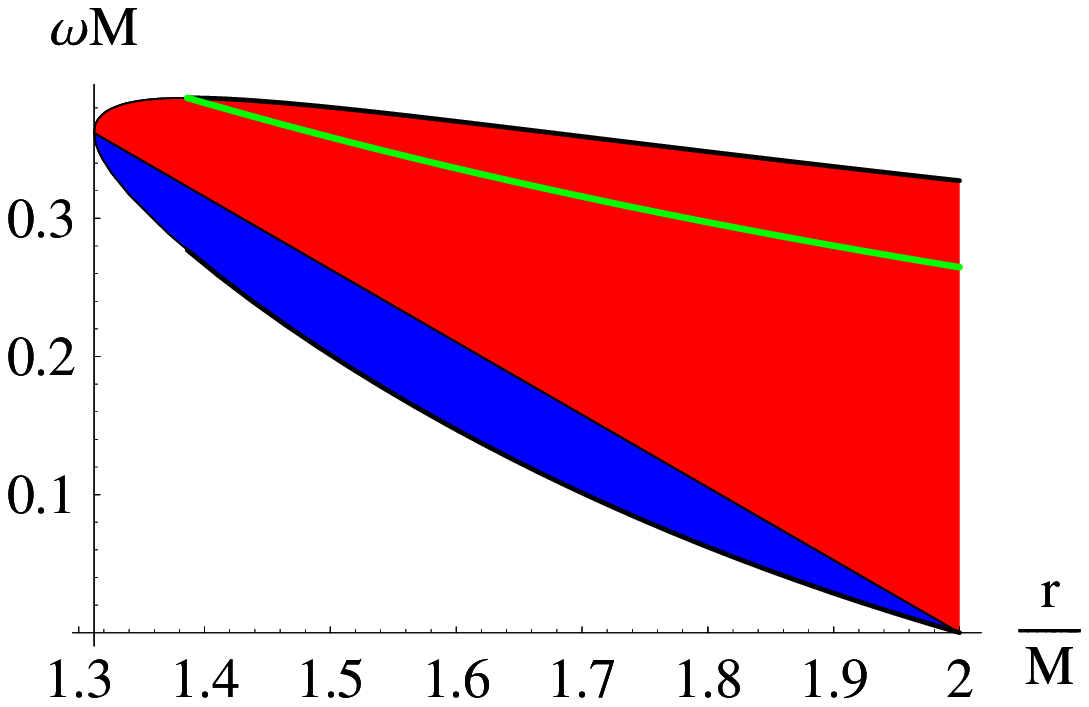} \ \ \ \
\includegraphics[width=63mm]{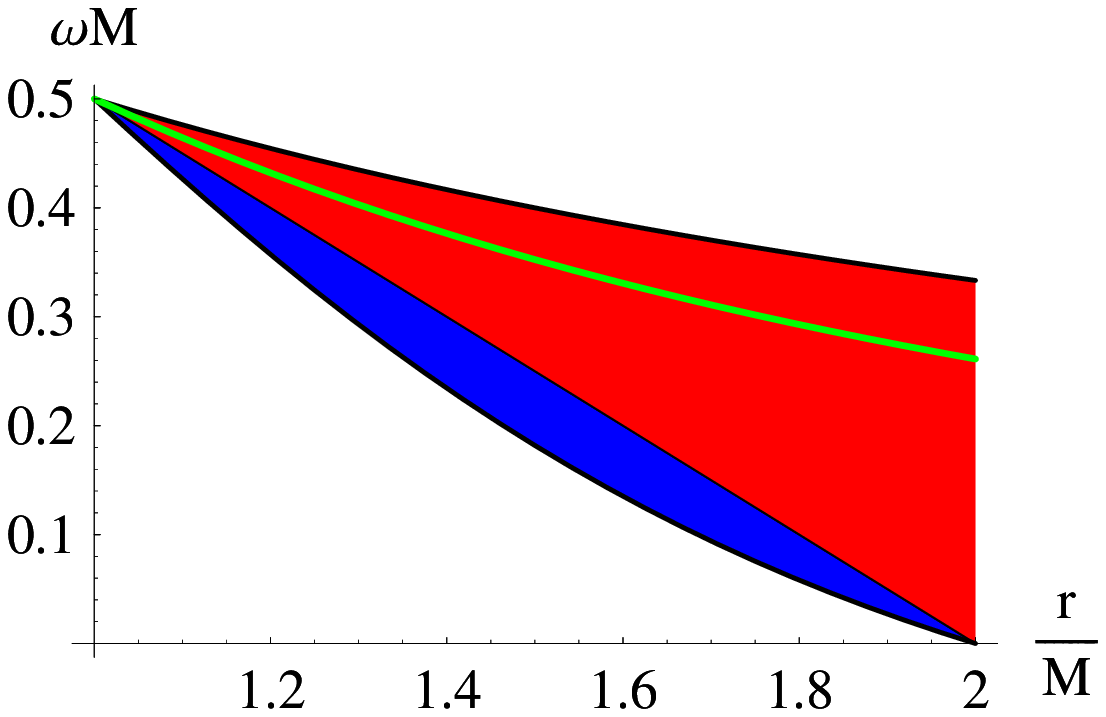}\\
\hspace{0mm}(c) \hspace{61mm} (d) \\
\caption{Domains of possible values of the angular velocities of particles
in the ergosphere for $\theta = \pi/2$ with
(a)~$a/M=0.5$, (b) $a/M=1/\sqrt{2} $, (c) $a/M=0.95$, (d) $a/M=1$:
the red area corresponds to the angular velocity of positive-energy particles
and the blue area corresponds to the angular velocity of
negative energy particles.}
\label{FigOmeg}
\end{figure}

    Substituting the components of metric~(\ref{Kerr}) in~(\ref{omN2}),
we obtain the limit values for the angular velocity of a Kerr rotating
black hole:
    \begin{equation}
\Omega_{1,2} = \frac{2 M r a \sin \theta \mp \rho^2 \sqrt{\Delta}}
{\sin \theta \, \Sigma^2} = \omega \mp \frac{ \rho^2 \sqrt{\Delta}}
{\sin \theta \, \Sigma^2}.
\label{OmK1}
\end{equation}
    From the geodesic equations for the angular velocity of
freely moving particles, we obtain
    \begin{equation}
\frac{d \varphi }{d t} = \frac{2 M r a E +
\frac{S J}{\sin^2\! \theta} }{\displaystyle
\Sigma^{\mathstrut 2} E - 2 M r a J } =
\omega + \frac{ \Delta \rho^4 J }{\sin^2\! \theta \Sigma^2
( \Sigma^2 E - 2 M r a J )} .
\label{OmK2}
\end{equation}
    From restriction~(\ref{lHmdd}) previously found for particles with
negative energy, we obtain the estimate
    \begin{equation}
\frac{J}{E M} \ge \frac{\sin \theta}{- S(r)} \left[
2 r a \sin \theta + \frac{1}{M} \sqrt{ \Delta \left[
\rho^4 - \left( \frac{m^2}{E^2} \rho^2 + \frac{\Theta}{M^2 E^2} \right)
S(r) \right] } \right],
\label{OmK3}
\end{equation}
    and for particles with positive energy in the ergosphere,
we obtain the estimate
    \begin{equation}
\frac{J}{E M} \le \frac{\sin \theta}{- S(r)} \left[
2 r a \sin \theta - \frac{1}{M} \sqrt{ \Delta \left[
\rho^4 - \left( \frac{m^2}{E^2} \rho^2 + \frac{\Theta}{M^2 E^2} \right)
S(r) \right] } \right].
\label{OmK4}
\end{equation}
    Substituting the boundary values of expressions~(\ref{OmK3})
and (\ref{OmK4}) for $m=0$ and $\Theta = 0$ in~(\ref{OmK2}), we obtain
expressions for $\Omega_1$ and $\Omega_2$ (see formulas~(\ref{OmK1})).
    Taking into account that the angular velocity $d \varphi / d t $
according to~(\ref{OmK2}) increases as $J/(EM)$ on each of the continuity
intervals $ \left( -\infty,\, \Sigma^2/(2 M^2 r a) \right)$
and $ \left( \Sigma^2/(2 M^2 r a),\, +\infty \right) $, we obtain a partition
of the possible angular velocities in the ergosphere into two subdomains:
the lower for particles with negative energy and the upper for particles
with positive energy (see Fig.~\ref{FigOmeg}).

    Passing to the limit $ |J/(EM)| \to \infty $ in equality~(\ref{OmK2}),
we obtain expression~(\ref{omE0}).
    Therefore, the boundary of the areas for particles with positive and
negative energy is a line corresponding to the angular velocity of zero-energy
particles.
    In the case of rotation in the equatorial plane, this is a straight line.
    Therefore, the value of the angular velocity of a particle in
the ergosphere at a given value of the radial coordinate clearly indicates
the sign of the particle energy and the value of the ratio $J/(EM)$.
    Zero-energy particles are particles that move in the ergosphere with
the angular velocity determined by~(\ref{omE0}).
    They separate particles in the ergosphere with negative energy from
the particles with positive energy by the value of the angular velocity at
a given~$r$
    {\it Particles with negative energy are particles that rotate in
the ergosphere with an angular velocity less than velocity~(\ref{omE0})!}

    From formulas~(\ref{OmK1}), it is easy to obtain that
$$
\theta = \frac{\pi}{2} \ \ \Rightarrow \ \
\left. \frac{\partial \Omega_2}{\partial r} \right|_{r=2M} =
\frac{1-2A^2}{\left( 2+A^2 \right)^2},
$$
    and the tangent to the graph of the function $\Omega_2(r)$ at
the point with $r=2M$ (on the ergosphere boundary) is therefore horizontal at
$A=1/\sqrt{2}$, as seen in Fig.~\ref{FigOmeg}\,b.
    The light green line line in Figs.~\ref{FigOmeg}\,c and \ref{FigOmeg}\,d
corresponds to circular equatorial orbits of massive particles, and its limit
point on the boundary of possible angular velocities corresponds to a circular
photon orbit.
    Circular equatorial orbits in the ergosphere are only possible for
particles with positive energy and only if $A \ge 1/\sqrt{2}$.
    As is known~\cite{Chandrasekhar}, \cite{NovikovFrolov},
\cite{BardeenPressTeukolsky72}, the smallest possible value of the radius of
a circular equatorial orbit at a given $A$ corresponds to the circular orbit
of photons with $J>0$ and is equal to
$$
r_+ = 2 M \left[ 1 + \cos \left( \frac{2}{3}
\arccos \frac{-a}{M} \right) \right].
$$

\vspace{2mm}
{\section{\large The collision energy of particles in the center-of-mass
system}
\label{secX}}

    We find the energy $E_{\rm c.m.}$ in the center-of-mass system of two
colliding particles with the rest masses~$m_1$ and $m_2$ by squaring
the expression
    \begin{equation} \label{SCM}
\left( E_{\rm c.m.}, 0\,,0\,,0\, \right) = p^{\,i}_{(1)} + p^{\,i}_{(2)},
\end{equation}
    where $p^{\,i}_{(n)}$ is the four-momentum of the particles $(n=1,2)$.
    Because $p^{\,i}_{(n)} p_{(n)i}= m_n^2$, we have
    \begin{equation} \label{SCM2af}
E_{\rm c.m.}^{\,2} = m_1^2 + m_2^2 + 2 p^{\,i}_{(1)} p_{(2)i} .
\end{equation}
    For free-falling particles with the energies $E_1$ and $E_2$
(at infinity) and the angular momenta $J_1$ and $J_2$, we obtain
    \begin{eqnarray}
E_{\rm c.m.}^{\,2} = m_1^2 + m_2^2 - \frac{2}{\rho^2 } \sigma_{1 \theta} \sigma_{2 \theta}
\sqrt{ \Theta_1 \Theta_2 } +  \hspace{21mm}
\nonumber \\
+\, \frac{2}{\Delta \rho^2} \left[ E_1 E_2 \Sigma^2 - 2 M r a (E_1 J_2 + E_2 J_1 ) -
J_1 J_2 \frac{S}{\sin^2 \! \theta} - \sigma_{1 r} \sigma_{2 r} \sqrt{ R_1 R_2}
\right]
\label{Col3}
\end{eqnarray}
    from the geodesic equations.
    If the motion of the particles is codirected along the radial coordinate
($\sigma_{1 r} \sigma_{2 r} =1$), then in the limit $r \to r_H$,
resolving the uncertainty of the type $0/0$ in the last term in~(\ref{Col3}),
we obtain
    \begin{eqnarray}
E_{\rm c.m.}^{\,2} = m_1^2 + m_2^2
- \frac{2}{\rho^2 } \sigma_{1 \theta} \sigma_{2 \theta} \sqrt{ \Theta_{1 H} \Theta_{2 H}}
+ \left( m_1^2 + \frac{\Theta_{1 H}}{\rho_H^2}
\right) \frac{J_{2 H} - J_2}{J_{1 H} - J_1 } +
\nonumber \\
+\, \left( m_2^2 + \frac{\Theta_{2 H}}{\rho_H^2} \right)
\frac{J_{1 H} - J_1}{J_{2 H} - J_2 } +
\frac{\rho^2_H} { 4 M^2 r_H^2 \sin^2 \! \theta } \frac{( J_{1 H} J_2 - J_{2 H} J_1)^2}
{(J_{1 H} - J_1 ) (J_{2 H} - J_2 )}
\label{Col4}
\end{eqnarray}
    for collisions on the horizon.
    If one of the particles has the angular momentum projection $J =J_H$
(critical particle) and the second has $J \ne J_H$,
then the energy of collision on the horizon diverges.
    This was found in the paper of
Ba\~{n}ados-Silk-West~\cite{BanadosSilkWest09} for extreme rotating black holes.
    For nonextreme black holes in the neighborhood of the horizon,
the value $J =J_H$ is inadmissible, but with multiple collisions, generally
speaking, it is possible~\cite{GribPavlov2010,GribPavlov2011} that $J$
reaches values arbitrarily close to $J_H$ as $r \to r_H$.
    We can obtain an arbitrarily high collision energy if the particle acquires
an angular momentum projection that is large in absolute value but negative as
a result of multiple collisions or the influence of external
fields~\cite{GribPavlov2013,GribPavlov2012}, and this is possible with
a fixed value of the particle energy in accordance with
inequality~(\ref{lHmdd}).

    The energy of head-on collisions ($\sigma_{1 r} \sigma_{2 r} =-1$)
on the horizon diverges,
    \begin{equation}
E_{\rm c.m.}^{\,2} \sim \frac{4 a^2 }{\Delta \rho^2}
(J_{1H} - J_1) (J_{2H} - J_2) \to \infty , \ \ \ r \to r_H,
\label{Col6}
\end{equation}
    if $J_i  \ne J_{i H}$.
    This option to achieve ultrahigh collision energy is available for
particles moving along the geodesic of
a white hole~\cite{GribPavlov2015,GribPavlov2015b}.

    Formulas~(\ref{SCM})--(\ref{Col6}) hold for any of colliding particles
with both positive and negative energy.
    For all values of the particle energy, the energy in the center-of-mass
system always satisfies
$$
E_{\rm c.m.} \ge m_1 + m_2,
$$
    because the colliding particles in the center-of-mass system are moving
toward each other at a certain speed.
    Three of the above ways to achieve infinitely large collision energy can
also be realized for particles with a negative (zero) energy.

\begin{figure}[ht]
\centering
\includegraphics[width=90mm]{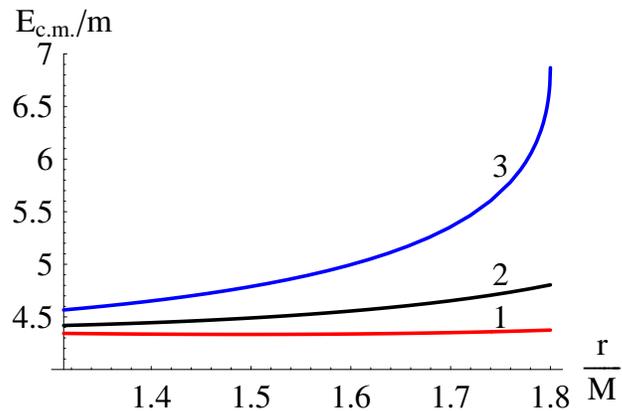} \ \ \ \
\caption{Collision energy of a particle with $E=m$ and $J=0$
and a second particle with $J=-16.487 m M $ and the energy $E_2=m$
(curve~{\bf 1}), $E_2=0$ (curve~{\bf 2}), or $E_2=-m$ (curve~{\bf 3}).}
\label{FigSt}
\end{figure}

    Calculations of the collision energy of two particles of equal mass $m$,
of which one falls from infinity into a black hole with $a=0.95M$ and
the other has positive, zero, or negative energy are shown in Fig.~\ref{FigSt}.
    At the selected value $J \approx -16.487mM $ of the angular momentum
projection, the radial component of the velocity of the negative-energy
particle is equal to zero at the point with $ r=1.8M $ for a black hole
with $a=0.95M$.
    Therefore, the collision energy is maximum here.
    The other particles fall with a nonzero radial velocity in
the range $ r \in (M, 1.8 M) $, and the collision energy is less.

{\bf Acknowledgements.}\,
    This research was supported by the Russian Foundation for Basic Research
(Grant No. 15-02-06818-a) and by
funds from a subsidy provided in the framework of government support of
Kazan Federal University for the purpose of increasing its competitiveness
among leading world science education centers.

\vspace{4mm}

\end{document}